\documentclass[twocolumn,aps,pra,showpacs,amsmath,amssymb,superscriptaddress]{revtex4}

\usepackage[usenames]{color}
\usepackage[dvips]{graphicx}

\begin{document}

\title{Ultracold Three-body Recombination in Two Dimensions}
\author{J. P. D'Incao}
\affiliation{JILA, University of Colorado and NIST, Boulder, Colorado 80309-0440, USA}
\affiliation{Department of Physics, Kansas State University, Manhattan, Kansas 66506, USA}
\author{Fatima Anis}
\affiliation{Department of Physics, Kansas State University, Manhattan, Kansas 66506, USA}
\author{B.D. Esry}
\affiliation{Department of Physics, Kansas State University, Manhattan, Kansas 66506, USA}

\begin{abstract}
We study three-body recombination in two dimensions for systems interacting via short-range two-body
interactions in the regime of large scattering lengths. Using the adiabatic hyperspherical representation, we 
derive semi-analytical formulas for three-body recombination in both weakly and deeply bound diatom states. 
Our results demonstrate the importance of long-range corrections to the three-body potentials by showing how they
alter the low-energy and scattering length dependence of the recombination rate for both bosonic and
fermionic systems, which exhibit suppressed recombination if compared to the three-dimensional case. 
We verify these results through numerical calculations of recombination for systems with finite-range 
interactions and supporting a few two-body bound states. We also study finite-range effects for the 
energies of the universal three-identical-bosons states and found a slow approach to universal 
predictions as a function of the scattering length.
\end{abstract}


\maketitle

\section{Introduction}

The advances in controlling interatomic interactions in ultracold gases through Feshbach resonances 
and the ability to confine these systems in anisotropic traps \cite{bloch2008RMP,chin2010RMP} have opened up 
ways to explore few-body systems in several new physical regimes. One such regime is obtained by strongly confining atoms in 
one dimension to produce an effective two-dimensional (2D) trap. Ultracold 2D gases have been the subject of intense theoretical and experimental explorations 
\cite{bloch2008RMP,petrov2000PRL,gorlitz2001PRL,rychtarik2004PRL,martiyanov2010PRL,rnacaglia2010PRL,dyke2011PRL,feld2011Nat,sommer2012PRL,
koschorreck2012Nat,zhang2012PRL} and have also been shown to display
several features relevant to condensed matter systems.

From the few-body perspective \cite{lapidus1982AJP,verhaar1984JPA,nielsen2001PR,petrov2001PRA,kanjilal2006PRA,liu2010PRB,brunch1979PRA,
nielsen1999FBS,platter2004FBS,hammer2004PRL,blume2005PRB,brodsky2006PRA,
kartavtsev2006PRA,lee2006PRA,bellotti2011JPB,bellotti2012PRA,bellotti2013JPB}, 
when interatomic interactions are strong, i.e., when the $s$-wave 2D scattering length, $a$, 
greatly exceeds the van der Waals length, $r_{\rm vdW}$, or any other short-range length scale, the system acquires universal 
properties that are manifested in both its bound and scattering properties. 
Universal few-body states have been studied for homonuclear and heteronuclear bosonic systems 
with strong $s$-wave interactions \cite{brunch1979PRA,nielsen1999FBS,platter2004FBS,hammer2004PRL,blume2005PRB,brodsky2006PRA,
kartavtsev2006PRA,lee2006PRA,bellotti2011JPB,bellotti2012PRA,bellotti2013JPB} 
and represent a novel class of states that might be accessible in experiments in 2D ultracold gases.
Their importance for many-body behavior is a question of much interest in the ultracold community.
More recently \cite{nishida2013PRL,volosniev2013ARX,gao2014ARX}, it has been shown that systems of three identical fermions 
near a $p$-wave resonance can display properties similar to the Efimov effect in three dimensions (3D) 
\cite{braaten2006PR,wang2013AAMOP}. 
Near a $p$-wave resonance, an infinity of universal 2D three-fermion states \cite{nishida2013PRL} can be formed, 
following a double exponential scaling,
even if two of the fermions cannot bind. This effect has also been shown to persist for heteronuclear three-body
systems \cite{moroz2014ARX} in 2D.
 
Despite the progress in understanding bound properties of few-body systems in 2D, some of their
scattering properties have yet to be understood. In particular, Refs.~\cite{helfrich2011PRA,ngampruetikorn2013EPL} have 
shown that three-body recombination vanishes at ultracold energies but its analytic behavior is not known. This is
in contrast to the 3D case, where recombination is constant at ultracold energies \cite{esry2001PRA}, but similar to 
one dimension (1D) \cite{mehta2007PRA}.
Three-body recombination is the process in which three free atoms collide to form a diatom and an atom, freeing  
enough kinetic energy to eject them from typical traps. 
Therefore, recombination is crucially important for ultracold 2D-gas experiments, and the understanding of 
the physics behind its suppression is of both practical and fundamental interest. In fact, a Wigner threshold law analysis for 2D recombination 
in the absence of resonant interactions leads to a constant value for 
recombination \cite{dincao2014ARX}, indicating that the strength of the interatomic interactions plays a fundamental role in determining 
the low-energy behavior of three-body recombination.

In this paper, we explore the scattering aspects of three-body systems in 2D. 
Using the adiabatic hyperspherical representation for zero-range two-body interactions \cite{nielsen2001PR},
we derive the asymptotic behavior of the three-body adiabatic potentials in order to explore the properties of 
recombination in 2D. We find that the long-range behavior of the three-body adiabatic potentials is responsible for the 
strong suppression in Refs.~\cite{helfrich2011PRA}. We determine semi-analytical formulas containing  
both the energy and the scattering length dependence for three-body recombination into weakly and deeply bound diatom states as well as 
the regime in which such low-energy results apply. We test the validity of our semi-analytical results through 
comparisons with numerical calculations using the adiabatic hyperspherical representation developed in Ref.~\cite{dincao2014ARX} 
for three-body systems with finite-range interatomic interactions. Using this methodology, we also study finite-range effects on the energies 
of three-boson bound states and find a very slow approach to their expected universal behavior in the limit $a\gg r_{\rm vdW}$ 
\cite{brunch1979PRA,nielsen1999FBS,platter2004FBS,hammer2004PRL,
blume2005PRB,brodsky2006PRA,kartavtsev2006PRA}. We derive an expression for the three-body energies in terms of the 
two-body effective range similar to the one derived in Ref.~\cite{helfrich2011PRA}, but for a regime not accessible in that work.

\section{Two- and Three-body physics in two dimensions}

\subsection{Two bodies}

The study of low-energy properties of two-body physics in 2D is facilitated by the restriction on the 
number of partial waves contributing to physical observables. For instance, scattering properties of two identical 
bosons, or of two distinguishable atoms, can be accurately described by the lowest angular momentum contribution, 
$m_{2b}=0$. In this case, the two-body scattering length, $a$, is the fundamental quantity that the various
properties of the system depends upon, and it is defined from the low energy expression 
for the $m_{2b}=0$ phaseshift \cite {lapidus1982AJP,verhaar1984JPA,nielsen2001PR,petrov2001PRA,kanjilal2006PRA}:
\begin{align}
\lim_{k\to 0 }\cot\delta_{m_{2b}=0} = \frac{2}{\pi}\left[\gamma_E + \ln(k_{2b}a/2)\right].\label{2bphaseshift}
\end{align}
Here, $\gamma_E\approx0.577216$ is Euler's constant, and $k_{2b}^2=2\mu_{2b}E$ is
the wave vector with $\mu_{2b}$ the two-body reduced mass and $E$ the two-body energy. 
Note that atomic units will be used throughout this paper unless otherwise stated.
The scattering phaseshift can be obtained by solving the two-body radial Schr\"odinger equation
\begin{eqnarray}
\left[-\frac{1}{2\mu_{2b}}\frac{d^2}{dr^2}+\frac{m_{2b}^2-1/4}{2\mu_{2b}r^2}+v(r)-E\right]f(r)=0,\label{2BEq}
\end{eqnarray}
with the appropriate set of asymptotic boundary conditions for scattering states \cite{lapidus1982AJP,verhaar1984JPA}.
In the above equation, $r$ is the interparticle distance; $v(r)$, the interatomic interaction; and $f(r)$, the corresponding radial 
wavefunction. Note that in 2D the scattering length $a$ in Eq.~(\ref{2bphaseshift}) is always positive.

Like the 3D problem, the large values of $a$ and the long wavelengths characteristic of ultracold energies lead to a
set of universal properties for 2D ultracold gases.
The simplest manifestation of such universal properties is the existence of weakly bound two-body states 
whose energies are determined irrespective of the details of the interatomic interactions (see, for instance, 
Ref. ~\cite{kanjilal2006PRA}). For the $m_{2b}=0$ case, for instance, assuming a zero-range pseudo-potential, one 
can determine the binding energy to be
\begin{align}
E_{2b} = \frac{2e^{-2\gamma_{E}}}{\mu_{2b}a^2}.\label{zrEb}
\end{align}
Therefore, $E_{2b}$ depends only on $a$ and not on any other property of the underlying interaction. 
We will show that for realistic systems this result is only valid when $a$ greatly exceeds the characteristic range 
of the interaction. These results, Eqs.~(\ref{2bphaseshift}) and (\ref{zrEb}), are, of course, simply the effective range
expansion long known in scattering (see for instance \cite{verhaar1984JPA}), but have found a particularly clear physical manifestation
in ultracold systems. 

\begin{figure}
\includegraphics[width=3.3in]{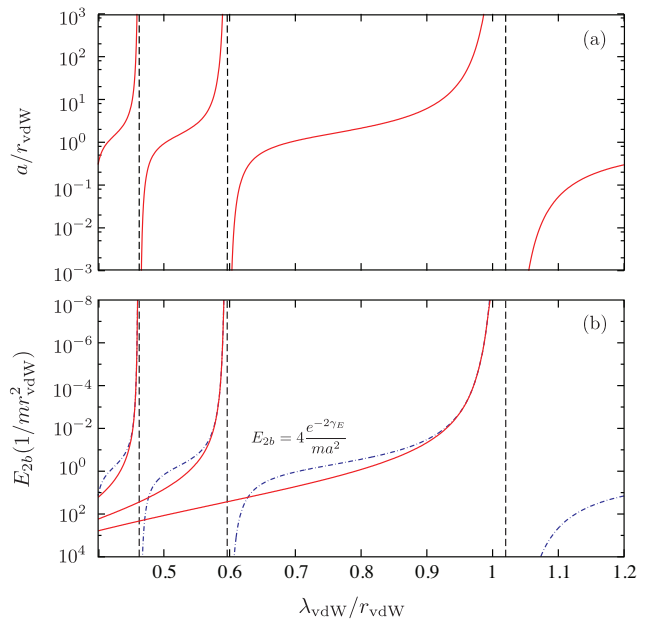}
\caption{(Color online) (a) 2D two-body scattering length, $a$, and (b) binding energy, $E_{2b}$, 
as a function of $\lambda_{\rm vdW}$. As $\lambda_{\rm vdW}$ decreases, $a$ [solid lines in (a)] 
goes from 0 to $+\infty$ every time a new bound state is formed, indicated in the figure by the vertical 
dashed lines. In (b) we compare the numerical values for $E_{2b}$ (solid lines) with the ones obtained from 
Eq.~(\ref{zrEb}) (dot-dashed lines).}
\label{ScaLEb2D}
\end{figure}

In Fig.~\ref{ScaLEb2D} we illustrate the above properties of $m_{2b}=0$ two-body 2D systems by solving 
Eq.~(\ref{2BEq}) for two identical bosons ---$\mu_{2b}=m/2$ where $m$ is the atomic mass--- interacting via
the Lennard-Jones potential 
\begin{eqnarray}
v(r)=-\frac{C_6}{r^6}\left(1-\frac{\lambda_{\rm vdW}^6}{r^6}\right).\label{vdWPot}
\end{eqnarray}
Here, $C_{6}$ is the dispersion coefficient and $\lambda_{\rm vdW}$ is a parameter used to produce the desired variations of $a$.
Note that in Fig.~\ref{ScaLEb2D}, and in what follows, we present results in van der Waals units, i.e., length is given in units of the van 
der Waals length, $r_{\rm vdW}=(2\mu_{2b}C_6)^{1/4}/2$, and energy in units of $1/m r_{\rm vdW}^2$. This eliminates any explicit dependence
on $C_{6}$. In Figs.~\ref{ScaLEb2D}(a) and (b), we show the 2D scattering length and binding energy, respectively, 
as a function of $\lambda_{\rm vdW}$. 
As $\lambda_{\rm vdW}$ decreases, the repulsion in Eq.~(\ref{vdWPot}) for $r\le\lambda_{\rm vdW}$ weakens, 
allowing the attractive term to become increasingly dominant. (The minimum of the potential in Eq.~(\ref{vdWPot}) occurs at
$r=2^{1/6}\lambda_{\rm vdW}$ and has the value $-4r_{\rm vdW}^4/m\lambda_{\rm vdW}^6$.)
In the process, multiple bound states form, each of which causes $a$ to diverge from 0 to 
$+\infty$ (see the vertical dashed lines in Fig.~\ref{ScaLEb2D}).

In Fig.~\ref{ScaLEb2D}(b) we compare the exact values of $E_{2b}$ obtained using the potential in Eq.~(\ref{vdWPot})
with the approximate one from Eq.~(\ref{zrEb}) showing, as expected, that the zero-range result is only
a good approximation for $a\gg r_{\rm vdW}$. 
In fact, we found numerically that for $a=10.04r_{\rm vdW}$ the agreement is on the $20\%$ level while
agreement to $1\%$ is only achieved for $a\gtrsim 100r_{\rm vdW}$. 
This quantifies the regime in which one should expect the universality as expressed in Eq.~(\ref{zrEb}) to 
be valid. 
To improve the agreement, recent work \cite{helfrich2011PRA,petrov2004PRL,wang2011PRA} has retained the next term in the expansion
in Eq.~(\ref{2bphaseshift}) to include the effective range $r_e$, thus incorporating at least some information about short-range physics.
When applied to observables like $E_{2b}$, this approach extends the concept of universality to dependence on $a$ and $r_e$.
(We perform an analysis of finite-range corrections in Sec. \ref{FiniteRange}.)

\subsection{Three bodies}

As a natural extension of the two-body analysis above, the 2D universality in three-body systems with large scattering lengths 
has been discussed in several recent 
works \cite{brunch1979PRA,nielsen1999FBS,platter2004FBS,hammer2004PRL,blume2005PRB,brodsky2006PRA,
kartavtsev2006PRA,lee2006PRA,bellotti2011JPB,bellotti2012PRA,bellotti2013JPB}.
In these studies, it has been shown that since there is no Efimov effect, no additional three-body 
parameter is required to determine low-energy three-body properties, i.e., the three-body physics is solely determined 
from two-body parameters. Nevertheless, on the three-body level, the 2D problem becomes more complex 
than the two-body problem due to the increase of the number of degrees of freedom. 

\subsubsection{Adiabatic Hyperspherical representation}

Here, we will briefly outline the main features of the 
adiabatic hyperspherical representation for the 2D three-body problem (details can be found in 
Refs.~\cite{nielsen2001PR,kartavtsev2006PRA,dincao2014ARX}) and discuss some of the universal properties of 
the system. Note that all results in this work apply to systems with three equal masses.

In the adiabatic hyperspherical representation, the hyperradius $R$ gives the overall size of system while all other 
degrees of freedom are described in terms of the set of hyperangles $\Omega$.
The total wave function is expanded in terms of the orthonormal channel functions
$\Phi_{\nu}(R;\Omega)$, 
\begin{equation}
\Psi(R,\Omega)=\frac{1}{R^{3/2}}\sum_{\nu}F_{\nu}(R)\Phi_{\nu}(R;\Omega),
\label{chfun}
\end{equation}
\noindent 
where $F_{\nu}(R)$ is the hyperradial wave function and $\nu$ represents
all quantum numbers necessary to specify each channel. The channel functions 
$\Phi_{\nu}(R;\Omega)$ are the eigenstates of the adiabatic equation
\begin{equation}
H_{\rm ad}(R,\Omega)\Phi_{\nu}(R;\Omega)
=U_{\nu}(R)\Phi_{\nu}(R;\Omega),\label{poteq}
\end{equation}
\noindent
whose eigenvalues, $U_{\nu}(R)$, are the three-body potentials from which the hyperradial motion is determined. 
Equation (\ref{poteq}) is solved for fixed values of $R$ with the appropriate set of boundary conditions 
\cite{nielsen2001PR,kartavtsev2006PRA,dincao2014ARX}, depending upon the definition of the hyperangles. 
For the present study, we use the democratic definition of the hyperangles \cite{dincao2014ARX,johnson1983JCP}
to facilitate imposing the identical particle symmetries \cite{dincao2014ARX}.

In the adiabatic equation, $H_{\rm ad}$ is the adiabatic Hamiltonian given by
\begin{eqnarray}
H_{\rm ad}(R,\Omega)=\frac{\Lambda^2(\Omega)+3/4}{2\mu R^2}+V(R,\Omega),\label{Had}
\end{eqnarray}
where $\mu=m/\sqrt{3}$ is the three-body reduced mass for atoms with identical masses $m$. Therefore, $H_{\rm ad}$ contains 
the grand angular momentum $\Lambda^2(\Omega)$, i.e., the hyperangular part of the kinetic energy,
as well as all the interparticle interactions via $V(R,\Omega)$. Here, we assumed $V(R,\Omega)$ to be 
a pairwise sum of the form
\begin{equation}
V(R,\Omega)=v(r_{12})+v(r_{23})+v(r_{31}), \label{Int}
\end{equation} 
where the interparticle distances $r_{ij}$ are given in terms of the hyperspherical
coordinates \cite{dincao2014ARX}. 

\begin{figure}[htbp]
\includegraphics[width=3.3in]{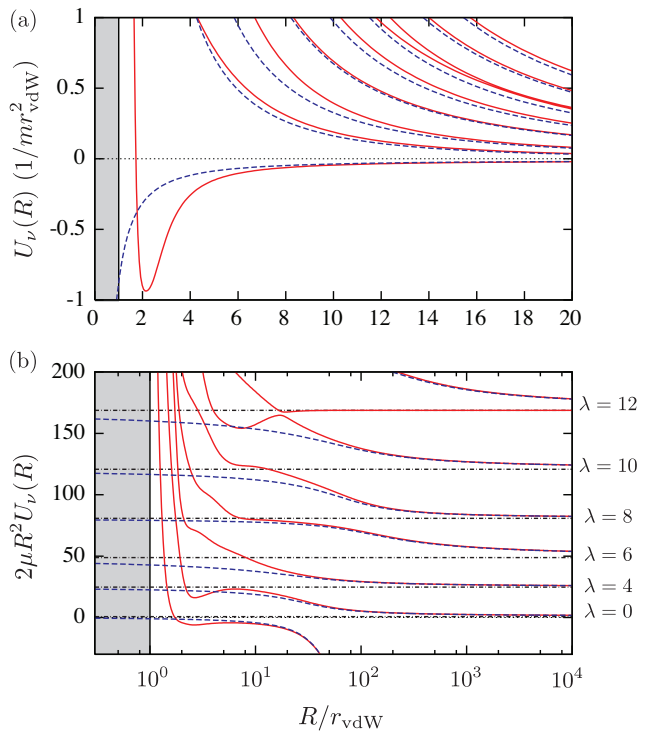}
\caption{(Color online) (a) Hyperspherical three-body potentials for $0^+_s$ $BBB$ systems with $a=10.04r_{\rm vdW}$. Results were
obtained using the finite-range interaction from Eq.~(\ref{vdWPot}) (red solid lines) and a zero-range model [Eq.~(\ref{TransEq})] 
(blue dashed lines). For this calculation, the lowest potential represents an atom-diatom channel, converging asymptotically ($R\gg a$) to 
the diatom energy $-E_{2b}$. All other potentials describe collisions between three free atoms, i.e., they represent three-body continuum channels
whose asymptotic behavior is described by $[\lambda(\lambda+2)+3/4]/2\mu R^2$. (b) Same as (a) but multiplied by $2\mu R^2$ in order to
emphasize the symmetry-allowed values for $\lambda$ (horizontal dash-dotted lines).}
\label{Pot_LJZR}
\end{figure}

Solving the adiabatic equation [Eq.~(\ref{poteq})] is the main task in the
adiabatic hyperspherical approach. In fact, once $\Phi_\nu(R;\Omega)$ and $U_\nu(R)$ are obtained, 
the problem becomes similar to the one in Eq.~(\ref{2BEq}) for two bodies. That is, one then has to solve
the hyperradial Schr\"odinger equation,  
\begin{multline}
\left[-\frac{1}{2\mu}\frac{d^2}{dR^2}+U_{\nu}(R)\right]F_{\nu}(R) \\
-\frac{1}{2\mu}\sum_{\nu'}
W_{\nu\nu'}(R)F_{\nu'}(R)
=EF_{\nu}(R),\label{radeq}
\end{multline}
\noindent
describing the hyperradial motion of the three-body system under the influence of the {\em effective} potentials $U_{\nu}(R)-W_{\nu\nu}(R)/2\mu$. 
The main difference from Eq.~(\ref{2BEq}) is the presence of nonadiabatic couplings $W_{\nu\nu'}(R)$ 
\cite{nielsen2001PR,kartavtsev2006PRA,dincao2014ARX}.
While Eq.~(\ref{radeq}) is exact when all channels are included, in practice the number of channels must be truncated,
but can be increased until the desired accuracy is achieved.

In Fig. \ref{Pot_LJZR} we show the three-body potentials obtained using the two-body interaction model
from Eq.~(\ref{vdWPot}) in the framework developed in Ref.~\cite{dincao2014ARX}. These results are for three identical bosons, $BBB$, with symmetry
$|M|^{\pi}_r=0^+_s$, where $M$ is the total orbital angular momentum, and $\pi$ is the overall parity, and $r$ is the quantum number that specifies the 
symmetric ($r=s$) and anti-symmetric ($r=a$) solutions with respect to the reflection $x\rightarrow-x$ \cite{dincao2014ARX}. For this example, 
we chose $\lambda_{\rm vdW}\approx0.923r_{\rm vdW}$, which produces 
$a=10.04r_{\rm vdW}$ while supporting a single $m_{2b}=0$ bound state.
Figure \ref{Pot_LJZR}(a) shows the two classes of three-body channels that, at distances $R\gg a$, represent atom-diatom collisions 
(lowest potential) and collisions between three free atoms (all other potentials). Their leading
order behavior is given, respectively, by
\begin{align} 
&U_{\nu}(R) 
\underset{R\gg a}{\longrightarrow}
 -E_{2b}+\frac{m_{AD}^2-1/4}{2\mu R^2}, \label{3BAD}
\end{align}
where $m_{AD}$ is the relative angular momentum between atom and diatom, satisfying $M=m_{2b}+m_{AD}$,
and by
\begin{align} 
&U_{\nu}(R) 
\underset{R\gg a}{\longrightarrow}
\frac{\lambda(\lambda+2)+3/4}{2\mu R^2}. \label{3BCont}
\end{align}
Here, $\lambda$ is the hyperangular momentum quantum number (a non-negative integer) determined
from the symmetry of the problem \cite{dincao2014ARX}. Figure \ref{Pot_LJZR}(b) shows the potentials 
from Fig.~\ref{Pot_LJZR}(a) multiplied by $2\mu R^2$ to emphasize their asymptotic approach to Eqs.~(\ref{3BAD}) and 
(\ref{3BCont}) as well as the allowed values of $\lambda$. 

\subsubsection{Zero-range model}

In Fig.~\ref{Pot_LJZR} we also show the results for the $BBB$ three-body potentials assuming a zero-range 
model for the interatomic interactions. As expected, the agreement between finite- and zero-range results
improves as $R$ increases and the details of the interatomic interactions become irrelevant.
(In Fig.~\ref{Pot_LJZR}, agreement between finite- and zero-range results are noticeable for 
$R\gtrsim10r_{\rm rvdW}$.)

For zero-range interactions, the adiabatic potentials are obtained by writing the potential as
\begin{eqnarray}
U_{\nu}(R)=\frac{s^2_{\nu}(R)-1/4}{2\mu R^2}\label{sUR}
\end{eqnarray}
where $s_{\nu}=2\xi+M+1$ and $\xi$ is determined from the transcendental equation
\cite{nielsen2001PR,kartavtsev2006PRA} (for equal mass systems)
\begin{widetext}
\begin{multline}
\left[\cos(\pi\xi)+\frac{\sin(\pi\xi)}{\pi}\left(\psi_{\Gamma}(\xi+1+M)+\psi_{\Gamma}(\xi+1)+
2\gamma_{E}-2\ln\left(\frac{R}{da}\right)
\right)\right]A_{1}
\\
+\frac{\Gamma(\xi+1+M)}{M !\Gamma(\xi+1)}F(-\xi,\xi+M+1;1+M;1/4)(-1/2)^{M}(A_{2}+A_{3})=0.\label{TransEq}
\end{multline}
\end{widetext}
In this equation, $d=3^{1/4}/2^{1/2}$ and $A_1=A_2=A_3$ are the coefficients for the Fadeev components \cite{nielsen2001PR} for three 
identical bosons. (For a system of two dissimilar ---but equal mass--- bosons, $BBB'$, or fermions, $FFF'$, one needs to set $A_1=A_2$, $A_3=0$ and $A_1=-A_2$, $A_3=0$,
respectively, assuming the identical particles do not interact.) 
Note that the values of $\lambda=s-1=2\xi+M$ in Eq.~(\ref{3BCont}) can be obtained by solving Eq.~(\ref{TransEq}) for $R\gg a$.

\section{Threshold behavior of three-body recombination}

As mentioned above, most of the recent work on 2D three-body physics has been focused on the bound properties of the
system, and important questions concerning the low-energy three-body scattering properties remain open. 
One particularly important process is three-body recombination, 
\begin{eqnarray}
X+X+X\rightarrow X_2+X+E_{2b}.
\end{eqnarray}
This is a major atom-loss mechanism in ultracold gases since its final products can have large kinetic energy, 
of the order of the binding energy of the diatom $X_{2}$, and are thus lost from typical traps. 
The findings of Refs.~\cite{helfrich2011PRA,ngampruetikorn2013EPL} point to a greater stability of ultracold 2D gases against 
recombination if compared with the 3D case. Whereas the 3D recombination rate $K_{3}$ for three identical bosons is constant at low 
energies \cite{fedichev1996PRL,esry1999PRL,nielsen1999PRL}, 2D recombination was found to vanish in this regime. 
However, a more physical interpretation of this important result is still lacking since a simple threshold analysis, such as
the one in Ref.~\cite{dincao2014ARX}, is incapable of explaining
it. In this section, we seek such an interpretation using a WKB approach
\cite{dincao2005PRL} and present semi-analytical results for both the scattering length and energy dependence of 
recombination using a zero-range interaction model. Finally, we compare these results with fully numerical finite-range calculations. 

It is well known that the low-energy dependence of scattering observables is controlled by 
the asymptotic form of the potential describing either the initial or final channels ---the initial channel for exothermic
(superelastic) collisions or the final channel for endothermic (inelastic) collisions.
Three-body recombination is no different, although calculating recombination normally requires the inclusion of
a large number of initial continuum channels [Eq.~(\ref{3BCont})], making the calculations extremely challenging. 
Fortunately, at ultracold energies, the lowest continuum channel, characterized by $\lambda=\lambda_{\rm min}$, 
provides the dominant contribution to recombination and allows one to derive analytical formulas for its energy and scattering length
dependence. 

Nevertheless, a simple WKB analysis \cite{dincao2014ARX} assuming the asymptotic potential in Eq.~(\ref{3BCont})
leads to $K_3\propto k^{2\lambda}$ ($k^2=2\mu E$). In otherwords, $K_3$ is constant for three identical bosons ($\lambda=\lambda_{\rm min}=0$) 
as $k\rightarrow0$ and is, therefore, in clear contradiction to the results of Ref.~\cite{helfrich2011PRA}. 
The same WKB analysis applied in 1D \cite{mehta2007PRA} and 3D \cite{dincao2005PRL} gives the correct results,
suggesting that the assumption of purely short-range corrections to Eq.~(\ref{3BCont}) ---and not
the method itself--- is to blame. Indeed, Eq.~(\ref{TransEq}) shows that the potential is a function of $\ln R/a$
and is thus likely to have important, relatively long-ranged, corrections. Note that this dependence on $\ln R/a$
rather than $R/a$ is peculiar to 2D.

\subsection{Asymptotic corrections to three-body potential}

In order to derive the long-range corrections to the potential in Eq.~(\ref{3BCont}), 
we introduce $R$ dependence into $\lambda$ and re-write Eq.~(\ref{3BCont}) as
\begin{align}
U_{\nu}(R) = \frac{\tilde\lambda({R})(\tilde\lambda({R})+2)+3/4}{2\mu R^2}.\label{UR}
\end{align}  
Since we seek the asymptotic behavior of $\tilde\lambda(R)$, we use the relation 
$\tilde\lambda(R)=2\xi(R)+M$ [obtained by equating Eqs.~(\ref{sUR})
and (\ref{UR})] and the fact that $\xi(R)=\xi(\ln R/da)$ from
Eq.~(\ref{TransEq}) to write its asymptotic expansion ($R\gg a$) as
\begin{align}
\tilde\lambda({R}) = \lambda+\sum_{n=1}^{\infty}\frac{c_{n}}{[\ln({R/da})]^n},\label{lambdaR}
\end{align}
with the coefficients $c_n$ determined by substitution into Eq.~(\ref{TransEq}). 
Equation (\ref{lambdaR}) allows us to write the asymptotic behavior of $U(R)$ as,
\begin{align}
&U_{\nu}(R) = \frac{\lambda(\lambda+2)+3/4}{2\mu R^2}+\frac{1}{2\mu R^2}\left[\frac{2(\lambda+1)c_{1}}{\ln({R/da})}\right.\nonumber\\
&\left.+\frac{c_1^2+2(\lambda+1)c_{2}}{\ln({R/da})^2}+\frac{2c_1c_2+2(\lambda+1)c_{3}}{\ln({R/da})^3}+\cdots\right].\label{URS}
\end{align}
The coefficients $c_1$, $c_2$, and $c_3$ depend on $\lambda$ and $M$, as well as on the permutation symmetry.
These constants can be derived analytically; their expressions,
however, are too cumbersome to include here. Therefore, in 
Table \ref{scaletab} we only list their numerical values for the lowest few values of $M$, $\lambda=\lambda_{\min}$, and 
different permutation symmetries. 

Our analysis thus shows that the leading-order correction goes as $1/[R^2\ln(R/da)]$ ---which is, strictly speaking,
a short-range potential. Nevertheless, we will show that it dramatically modifies the threshold behavior of recombination.
We also note that, although nonadiabatic correction 
to the potential $U(R)$ are typically important, the leading-order nonadiabatic 
corrections to the three-body continuum channels is proportional to $1/[R^2\ln(R/da)^4]$ \cite{kartavtsev2006PRA}, and we can
neglect such terms in our analysis.

\begin{table}
\caption{Coefficients for the long-range corrections to the three-body potential in Eq.~(\ref{URS}) for the lowest few values of $|M|$ and
$\lambda=\lambda_{\min}$ for different permutation symmetries. We also list the corresponding values for $\gamma$ in Eqs.~(\ref{K3w}) and (\ref{K3d}).}
\begin{ruledtabular}
\begin{tabular}{ccccccc}
         & $|M|_r^{\pi}$ & $\lambda_{\rm min}$ & $c_{1}$ & $c_{2}$ & $c_{3}$ &  $\gamma$\\
\hline
$BBB$  & $0^+_s$  & 0   & 3   & --0.86305 &  6.44605 &  2.8239\\
           & $1^-_s$  & 3   & 2.25 &  2.54857  &  2.04571 &  2.7491 \\
           & $2^+_s$  & 2   & 1.5   & 1.57962 &  1.42382 &  2.2018\\
$BBB'$ & $0^+_s$  & 0   & 2   & --0.28768 &  1.41866 &  2.4862\\
           & $1^-_s$  & 1   & 0.5   & 0.16096 & --0.04944 &  3.0144\\
$FFF'$ & $0^+_s$  & 2   & 1.5   & 1.01712 &  --0.00943 & 3.2314\\
         & $1^-_s$    & 1   & 1.5   &  1.01712 &  0.47603 &  2.1502
\end{tabular}
\end{ruledtabular}
\label{scaletab}
\end{table}

With Eq.~(\ref{URS}), it is now straightforward to derive the effect of the corrections to $U_\nu(R)$ 
on the low-energy behavior of recombination using our WKB approach \cite{dincao2005PRL}. 
The three-body recombination rate in 2D is given by
\begin{align}
K_{3} =n!\frac{4\pi}{\mu} \sum_{fi}\frac{|T_{fi}|^2}{k^{2}},
\label{tmateq}
\end{align}
where $n$ is the number of identical particles and $T_{fi}$ is the $T$-matrix element between initial, $i$, and final, $f$, states. 
For recombination, the initial states are the three-body continuum states described asymptotically ($R\gg a$) by the potentials in 
Eq.~(\ref{URS}). For large values of $a$, we classify the possible final states as weakly or deeply bound atom-diatom 
states and analyze recombination for each case separately. As we will see, each group recombines via different pathways, 
and their corresponding scattering length dependence differs substantially.

\subsection{Recombination into weakly bound states}

\begin{figure}[htbp]
\includegraphics[width=3.3in]{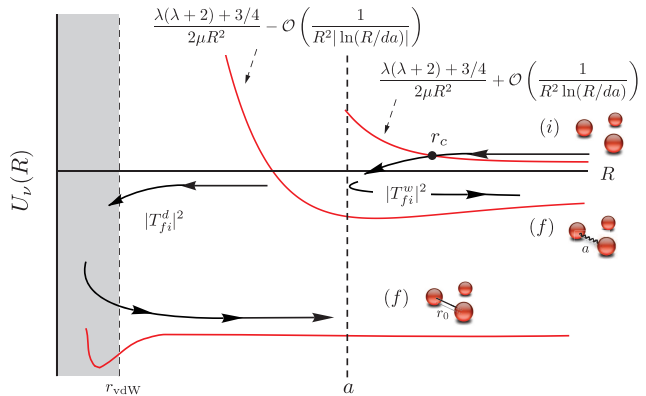}
\caption{(Color online). 
Schematic illustration of the relevant three-body potentials for three-body recombination and the corresponding collision pathways.
For recombination into weakly bound states, the initial free-atom state must tunnel from $R=r_c\gg a$ to $R\propto a$, 
where the inelastic transition to the final state, the weakly bound molecular channel (diatomic state connected by a wiggly line), occurs. 
For recombination into deeply bound diatoms, the pathway most likely to dominate includes an inelastic transition at $R\propto a$ 
to the weakly bound atom-diatom channel, but must subsequently tunnel from $R\propto a$ to $R\propto r_{\rm vdW}$, 
where an inelastic transition to the final deeply bound molecular channel (diatomic state bound by straight line) occurs.
}
\label{Pathways}
\end{figure}

For three-body recombination into weakly bound states, the energy and scattering length dependence can be determined \cite{dincao2005PRL} 
from the observation that inelastic transitions occur at distances proportional to $a$, i.e., when the initial free-atom
wave function has a substantial overlap with the final state atom-diatom wave function. 
At ultracold energies, these distances are much smaller than the classical turning point $r_c$.
The corresponding collision pathway is illustrated in Fig.~\ref{Pathways}, showing the initial tunneling from $R=r_{c}$ to $R\approx a$.
At this distance, the inelastic transition to the final weakly bound molecular channels occurs. Consequently, $|T_{fi}|^2$ from Eq.~(\ref{tmateq})
can be approximated by the WKB tunneling probability,
\begin{align}
&|T^{w}_{fi}|^2\approx \nonumber \\ 
& {\rm exp}\left[-2\int_{\alpha\times (da)}^{r_c}{\sqrt{2\mu\left(U_{\nu}(R)+\frac{1/4}{2\mu R^2}
-E\right)}dR}\right].
\label{TProbeq}
\end{align}
In the lower limit, $\alpha$ is an unknown constant on the order of 1 ---its precise value, however, does not affect either the energy or the
 scattering length dependence of $K_3$. 
Note that $U_\nu(R)$ is given by Eq.~(\ref{URS}) (also indicated in Fig.~\ref{Pathways}) and that
we have included the Langer correction \cite{berry1972RPP}. 

The integral in Eq.~(\ref{TProbeq}) cannot, however, be evaluated analytically. 
If, however, we neglect $E$, ic can be evaluated analytically, even
when several terms in $U_{\nu}(R)$ are retained. Since we seek only
the leading order correction, though, only the first two terms of Eq.~(\ref{URS}) are necessary.
To be consistent, we also expand the result and $r_c$, keeping leading-order corrections systematically.
Thus, with $r_c=(\lambda+1)/k$, we obtain a closed-form for Eq.~(\ref{TProbeq}) and
determine recombination into a weakly bound state to be,
\begin{align}
K^{w}_{3} = A^{w}_{\lambda}\frac{4\pi}{\mu}n!\frac{k^{2\lambda}a^{2\lambda+2}}{|\ln(ka/\gamma)|^{2c_1}},\label{K3w}
\end{align}
where $\gamma=(\lambda+1)/d$. Note that the $\alpha$ dependence in Eq.~(\ref{K3w}) appears only in the overall constant 
$A^{w}_{\lambda}$, whose value ---assumed to be universal--- can be determined by fitting Eq.~(\ref{K3w}) to numerically 
calculated $K_{3}^w$.

\begin{figure}[htbp]
\includegraphics[width=3.3in]{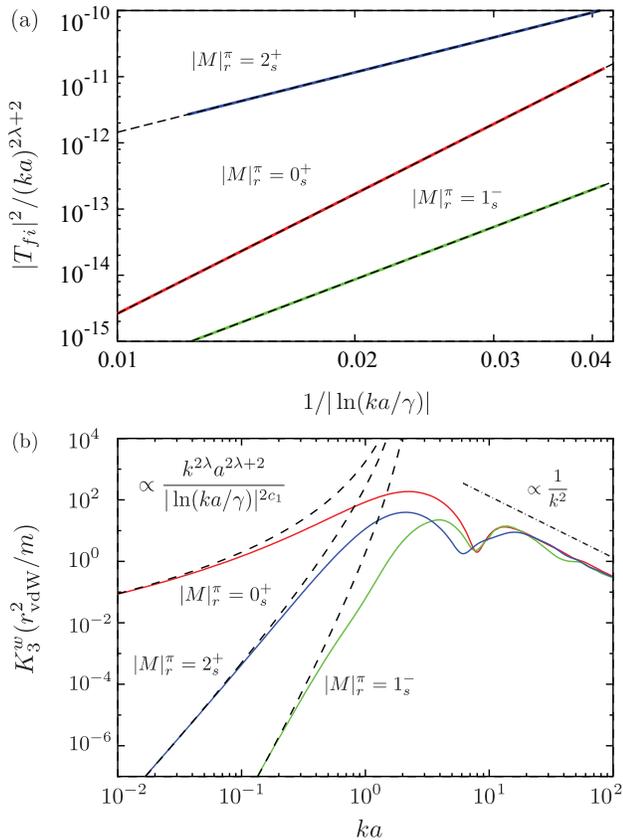}
\caption{(Color online) (a) Energy dependence of $|T_{fi}|^2$ for three-identical bosons with $|M|^\pi_s=0^+_s$, $1^-_s$ and $2^+_s$
obtained by solving Eq.~(\ref{TProbeq}) using the potentials $U(R)$ obtained from Eq.~(\ref{TransEq}) (thick solid lines) and clearly demonstrating 
the $1/|\ln(ka/\gamma)|^{2c_1}$ dependency on $|T_{fi}|^2$ (thin dashed lines). (b) corresponding numerical calculations for three-identical bosons 
recombination (solid lines) confirming the validity of Eq.~(\ref{K3w}) (dashed lines) for $ka\ll1$. 
Values for $\lambda$, $c_1$ and $\gamma$ are given in Table~\ref{scaletab}.\label{TWKBK3}}
\end{figure}

The threshold behavior expected for purely short-ranged corrections to Eq.~(\ref{3BCont}) ---the numerator of Eq.~(\ref{K3w}) 
\cite{dincao2014ARX}--- is clearly modified by the corrections in to the potential Eq.~(\ref{URS}).
As can be seen in Fig.~\ref{TWKBK3}, Eq.~(\ref{K3w}) fits the fully numerical evaluations of Eq.~(\ref{TProbeq}) with the full potential 
resulting from the exact solution of the zero-range-model transcendental equation [Eq.~(\ref{TransEq})]. These results were obtained by treating
$\gamma$ as a free parameter, however, to account for the higher-order potential terms and finite energy. 
Numerical values for $gamma$ are listed in Table \ref{scaletab}. (Numerical and fitted results differs in 0.1$\%$.) 

From Eq.~(\ref{K3w}) and Table \ref{scaletab}, it is apparent that both the energy and the scattering length dependence are strongly
affected by the logarithmic corrections to the asymptotic three-body potential shown in Eq.~(\ref{URS}).
For fixed $a$, the $1/|\ln(ka)|$ factor leads to the suppression of $K_3$ as $ka\rightarrow0$ found in Refs.~\cite{helfrich2011PRA,ngampruetikorn2013EPL}, 
while for fixed $k$ it shows the increase of $K_{3}$ as $a$ increases ($ka\ll1$).

In Fig.~\ref{TWKBK3}(b), we confirm Eq.~(\ref{K3w}) by comparing it to $K_{3}$ obtained from full numerical solutions of
Eqs.~(\ref{Had}) and (\ref{radeq}) for $BBB$ systems with $|M|^{\pi}_r=0^+_s, 1^-_s$, and $2^+_s$. 
The two-body interaction was taken from Eq.~(\ref{vdWPot}) and was chosen to have $a=10.04r_{\rm vdW}$
and a single $m_{2b}=0$ state. 
In order to determine $K_{3}$ numerically, we solved Eq.~(\ref{radeq}) using the methodology developed in 
Ref.~\cite{wang2011PRAb} up to distances comparable to $R=10^5r_{\rm vdW}$. Such large distances were required to ensure that the effect of the 
logarithmic terms in Eq.~(\ref{URS}) were negligible in order to properly satisfy the scattering boundary conditions.
In Appendix \ref{Match}, we present some of the details and a discussion of this particular issue.
The rates in Fig.~\ref{TWKBK3}(b) were obtained using up to 15 channels, giving three digits of accuracy in $K_3$ for energies up 
to $ka\approx1$ but fewer for $ka\approx10^2$. 

As expected, the numerical results agree well with the zero-range WKB prediction 
of Eq.~(\ref{K3w}) for $ka\ll1$. Similar to the 3D case, for $ka\gtrsim 1$ the system enters the regime 
where $|T_{fi}|^2$ approaches a constant value \cite{dincao2004PRL}, implying that $K_{3}\propto1/k^2$. 
This regime can be clearly seen in Fig.~\ref{TWKBK3}(b).

\subsection{Recombination into deeply bound states}

Three-body recombination into deeply bound diatoms proceeds through a different pathway than recombination into weakly bound
states. Inelastic transitions to deep states occur at distances comparable to the range of the interatomic interaction, 
in our case $r_{\rm vdW}$. Therefore, within our WKB approach, it will require knowledge of the three-body potentials for 
$r_{\rm vdW}\lesssim R\lesssim a$. Therefore, since that samples the potentials within the regime where $a\gg r_{\rm vdW}$,
we also expect universal behavior for recombination into deeply bound molecular states.
In this region, the three-body potential can be determined from Eq.~(\ref{URS}) by substituting $\ln(R/da)$ by $-|\ln(R/da)|$ which 
produces a repulsive barrier that prevents particles from approaching to short distances (see Fig.~\ref{Pathways}). 

The pathway most likely to dominate (also illustrated in Fig.~\ref{Pathways}) includes an inelastic transition at $R\approx a$ 
to the weakly bound atom-diatom channel with probability given by Eq.~(\ref{TProbeq}).
To reach $R\approx r_{\rm vdW}$, where the inelastic transition to the final deeply bound channel occurs, additional tunneling is required.
In our WKB approach, the probability to tunnel through the region $r_{\rm vdW}\lesssim R\lesssim a$ can be written as
\begin{align}
&|T^{d}_{fi}|^2\approx \nonumber \\ 
& {\rm exp}\left[-2\int_{r_{\epsilon}}^{\epsilon\times(da)}{\sqrt{2\mu\left(U_{\nu}(R)+\frac{1/4}{2\mu R^2}
-E\right)}dR}\right].
\label{TProbeqD}
\end{align}
Here, $\epsilon<1$ is an unknown constant whose precise value, similar to $\alpha$ in Eq.~(\ref{TProbeq}), does not affect the 
energy and scattering length dependence of $K_3$, and $r_{\epsilon}\propto r_{\rm vdW}$ is a short-range length scale that
can be determined by fitting numerical calculations. 

Proceeding with the same approximations as for the integral in Eq.~(\ref{TProbeq}), and
realizing that the total transition probability for recombination into deeply bound states is $|T_{fi}^{w}|^2|T_{fi}^{d}|^2$, 
we arrive at the expression for recombination 
\begin{align}
K^{d}_{3} =
A^{d}_{\lambda}\frac{4\pi}{\mu}n!\frac{k^{2\lambda}r_\epsilon^{2\lambda+2}}{|\ln(ka/\gamma)|^{2c_1}}\frac{1}{|\ln(a/r_{\epsilon})|^{2c_1}},\label{K3d}
\end{align}
where $A^{d}_{\lambda}$ is a nonuniversal constant, depending on the short-range physics
encapsulated in $r_{\epsilon}$. 
Notice that although the energy dependence of recombination into weakly 
and deeply bound states is the same, the scattering length dependence is different. 
In fact, the scattering dependence in $K_{3}^{d}$ implies a suppression of recombination into deeply bound states for
$a\gg r_{\rm vdW}$ ($ka\ll 1$) through the $1/|\ln (a/r_{\epsilon})|^{2c_1}$ term relative to $K_{3}^w$. 

We note that for the range of $a$ we explored, our numerical calculations do not exhibit a clear suppression of recombination into 
deeply bound states. Based on our calculations, we expect such suppression to occur only for $a>5000r_{\rm vdW}$, when 
the potential barrier in the weakly bound diatom channel for $r_{\rm vdW}<R<a$ is more evident. 
Nevertheless, this suppression is a feature of recombination in 2D that can allow for greater stability of 2D Bose gases 
in comparison to the 3D case, where $K_{3}^{d}\propto a^4$. 

\subsection{Recovering the naive threshold behavior}

Finally, we notice that when $a\rightarrow\infty$ or $a\rightarrow0$, the logarithmic terms in Eq.~(\ref{URS}) vanish, and the
potentials relevant to recombination are simply described by those for three free particles [Eq.~(\ref{3BCont})]. This indicates that for 
$a=\infty$ (and $a=0$) the energy dependence for recombination reduces to the one we derived in Ref.~\cite{dincao2014ARX}, i.e., 
$K_{3}\propto k^{2\lambda}r_{\rm vdW}^{2\lambda+2}$, predicting $K_{3}$ is constant for three identical bosons as $k\rightarrow0$. 

\begin{figure}[htbp]
\includegraphics[width=3.3in]{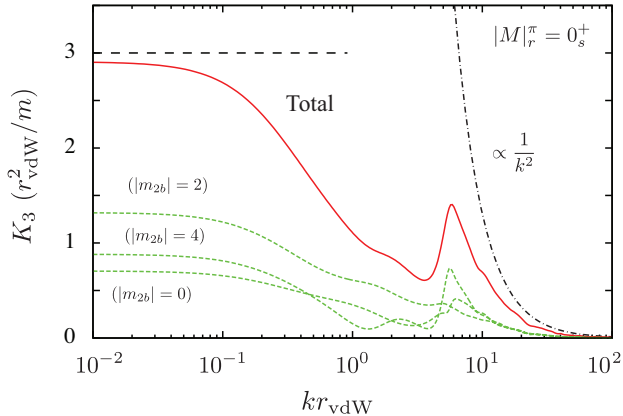}
\caption{(Color online) Three-body recombination for $0^+_s$ three identical bosons with $a\rightarrow\infty$.
For this calculation, the potential in Eq.~(\ref{vdWPot}) was adjusted to support three deeply bound two-body
states with $m_{2b}=0$, 2, and 4. Dashed lines indicate the partial recombination to each of these states;
and the solid line, the corresponding total rate. Note that, for $a\rightarrow\infty$ recombination is constant
in the limit of $k\rightarrow0$. \label{K3ScaL=Infty}}
\end{figure}

Figure~\ref{K3ScaL=Infty} shows the $a\rightarrow\infty$ behavior for the 
$BBB$-$|M|^{\pi}_r=0^+_s$ recombination rate, obtained for interactions supporting three two-body bound states 
[near the second pole in Fig.~\ref{ScaLEb2D}(a)] with $|m_{2b}|=0$, $2$ and $4$. 
We also note that for $a=\infty$, the threshold regime is characterized by values of $k$ in which $kr_{\rm vdW}\ll1$.  For values
of $kr_{\rm vdW}\gg1$, recombination enters the regime where $K_{3}^{d}\propto1/k^2$ \cite{dincao2004PRL}.

\section{Finite range corrections to three-body states in 2D} \label{FiniteRange}

It is well known that zero-range results are valid when the scattering length greatly exceeds all other
length scales in the system. 
Finite-range corrections to these results are generally assumed to be universal themselves, although 
establishing this as a fact is much harder due to the complexity of treating such corrections
\cite{helfrich2011PRA,petrov2004PRL,wang2011PRA}. 
According to our two-body calculations, for $a=10.04r_{\rm vdW}$ the zero-range binding energy [Eq.~(\ref{zrEb})] 
agrees with the numerical results within only 20$\%$, thus providing a first glimpse of the importance of 
finite-range corrections to the zero-range results.
Therefore, the natural questions are: (i) how small does $r_{\rm vdW}/a$ have to be so that the agreement between zero- and finite-range results is 
quantitatively obtained and (ii) how can finite-range corrections be incorporated in the zero-range model to improve the comparison? 

To give some sense of how the three-body universal limit is approached, we will focus on the three-body bound state energies,
calculating them with the two-body potential in Eq.~(\ref{vdWPot}) for increasing $a$ and comparing to the zero-range result.
As shown in Refs. \cite{brunch1979PRA,nielsen1999FBS,platter2004FBS,hammer2004PRL,blume2005PRB,brodsky2006PRA,
kartavtsev2006PRA}, for identical bosons in 2D there are always two three-body states associated with the weakly bound 
diatom state. Their energies are universally related to $E_{2b}$ as:
\begin{eqnarray}
E_{3b}^{(0)} \approx 16.523 E_{2b}~\mbox{~~~and~~~}~E_{3b}^{(1)} \approx 1.270 E_{2b}.\label{zrE3b}
\end{eqnarray}
(Note that $E_{3b}$ above is defined from the three-body breakup threshold. The three-body binding energy
is defined as $E_{3b}-E_{2b}$.) 
In Ref.~\cite{helfrich2011PRA}, corrections to these energies were obtained by keeping one more term in the effective 
range expansion of the two-body phaseshift in Eq.~(\ref{2bphaseshift}),
\begin{align}
\lim_{k\to 0 }\cot\delta_{m_{2b}=0} = \frac{2}{\pi}\left[\gamma_E + \ln(k_{2b}a/2)\right]+\frac{r_e^2k_{2b}^2}{2\pi}, \label{2bphaseshiftNew}
\end{align}
where $r_e$ is the effective range as defined in Ref.~\cite{verhaar1984JPA}. 

In Ref.~\cite{helfrich2011PRA}, a perturbative expansion in $r_e/a$ for the three-body energies
was obtained for the case in which $r_e^2<0$. In this case, the corrections to Eq.~(\ref{zrE3b})
 were found to be significant and thus imply a slow approach to universality. 
For our two-body interaction model [Eq.~(\ref{vdWPot})], however, the effective range correction $r_e^2k_{2b}^2/2\pi$ is always 
positive, a case for which Ref.~\cite{helfrich2011PRA} was not able to extract a similar perturbative expansion for the three-body energies. 
Similarly, our analysis below will be performed in terms of $r_e/a$. We note, however, that as we change $\lambda_{\rm vdW}$ 
in Eq.~(\ref{vdWPot}) both $a$ and $r_e$ change. [This allows us to relate such quantities and write $r_e\equiv r_{e}(a)$.]

\begin{figure}[htbp]
\includegraphics[width=3.3in]{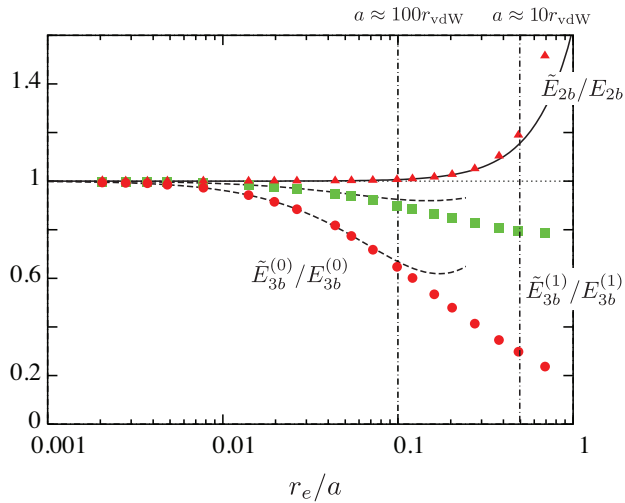}
\caption{(Color online) 
Comparison of numerical and analytical expressions for the energies of 2D two- and three-body bound states highlighting finite-range effects.
Filled triangles are the results for the two-body energies while filled circles
and squares are the results for the ground and excited three-body states, respectively.  Note that the results are 
normalized to the zero-range values from Eqs.~(\ref{zrEb}) and (\ref{zrE3b}).
The solid line corresponds to the two-body energy given by Eq.~(\ref{EbEffR}) while the dashed lines represent the three-body energies 
from Eqs.~(\ref{E3bG}) and (\ref{E3bE}).}
\label{E3bE2b}
\end{figure}

A first consequence of including the effective range term in Eq.~(\ref{2bphaseshift}) is that the two-body binding energy in Eq.~(\ref{zrEb}) 
can be corrected for $r_e/a\ll1$ to be
\begin{eqnarray}
\tilde E_{2b} \approx \frac{2e^{-2\gamma_{E}}}{\mu_{2b}a^2}\left[1+2e^{-2\gamma_{E}}\left(\frac{r_e}{a}\right)^2\right].\label{EbEffR} 
\end{eqnarray}
For $a=10.04r_{\rm vdW}$, the potential in Eq.~(\ref{vdWPot}) gives $r_e=4.89r_{\rm vdW}$. And although $r_e/a\approx0.49$ is relatively large, 
the above formula improves the agreement with the numerical results to $3\%$, as opposed to the $20\%$ deviation for the pure 
zero-range result from Eq.~(\ref{zrEb}). Therefore, the effective range correction to $E_{2b}$ greatly improves the
comparison with the finite-range results. For $r_{e}/a\approx0.1$ ($a\approx100r_{\rm vdW}$), the numerical value for $E_{2b}$ agrees to 
about $0.7\%$ with the pure zero-range result [Eq.~(\ref{zrEb})] and to about $0.01\%$ with the one from Eq.~(\ref{EbEffR}).
In Fig.~\ref{E3bE2b} we illustrate the agreement between our numerical values for $E_{2b}$
and the ones obtained using Eq.~(\ref{EbEffR}) as a function of $r_e/a$. Note that both results are 
normalized to the zero-range $E_{2b}$ obtained from Eq.~(\ref{zrEb}), or Eq.~(\ref{EbEffR}) with $r_e=0$.

Compared to the three-body energies ---also shown in Fig.~\ref{E3bE2b}--- the two-body energy converges to the zero-range
result relatively quickly. 
Here, too, $E_{3b}$ are normalized to the corresponding zero-range results in Eq.~(\ref{zrE3b}). 
While the numerical two-body energy for $a=10.04r_{\rm vdW}$ agrees at the $20\%$ level with the zero-range result, 
the ground and excited three-body energies deviate from the zero-range results [Eq.~(\ref{zrE3b})] by $70\%$ and $20\%$, respectively 
(see rightmost vertical dash-dotted line in Fig.~\ref{E3bE2b}). For $a\approx100r_{\rm vdW}$ (leftmost vertical dash-dotted line in 
Fig.~\ref{E3bE2b}), the deviation drops to $40\%$ and $10\%$, respectively.
The better agreement for the excited three-body state is consistent with the fact that its larger size would tend to minimize 
finite-range effects. By fitting our numerical results within the range $r_e/a<0.02$ to a perturbative expansion in $r_e/a$, 
we find that the formula
\begin{align}
&{\tilde E_{3b}^{(0)}}\approx16.52E_{2b}\left[1-1.76\left(\frac{r_{e}}{a}\right)^2\Big|\log\left(\frac{r_{e}}{a}\right)\Big|^{3.52}\right],\label{E3bG}\\
&{\tilde E_{3b}^{(1)}}\approx1.270E_{2b}\left[1-0.31\left(\frac{r_{e}}{a}\right)^2\Big|\log\left(\frac{r_{e}}{a}\right)\Big|^{3.82}\right],\label{E3bE}
\end{align}
describes our numerical results within this range to better than $1\%$. (it gradually improves for smaller values of $r_{e}/a$). 
This comparison is shown in Fig.~\ref{E3bE2b}. We have tried other forms for the expansion in Eqs.~(\ref{E3bG}) and
(\ref{E3bE}) and found, empirically, that including the logarithm term leads to more stable fits. Although empirical, this term serves as evidence 
of the slower approach to the zero-range results than for the two-body case.
Although deviations for the three-body energies for $r_e/a\gtrsim0.1$ can be observed,
it is not clear whether these deviations are themselves universal, i.e., they depend only on the effective range. 
In order to test the universality of Eqs.~(\ref{E3bG}) and (\ref{E3bE}) and higher-order corrections,
one would need to calculate the three-body energies using different finite-range 
two-body interaction models supporting different numbers of bound states ---a task beyond the scope of this study.

Finally, we note that in our numerical calculations of $r_e$, 
we found an approximate relation between $r_e$ and $a$ given by
$r_e/r_{\rm vdW}\approx2.063\ln(a/r_{\rm vdW})^{1.029}$ for the branch of $\lambda_{\rm vdW}$ [Eq.~(\ref{vdWPot})] giving a single $m_{2b}=0$ bound state.
We found this result by fitting our numerical calculations for $r_e/a<0.1$ and obtained an agreement under 0.1$\%$.
Calculations for the potential model $v(r)=D{\rm sech}^2(r/r_0)$ ---also supporting a single $m_{2b}=0$ bound state---, 
where $D$ is the potential depth and $r_{0}$ the characteristic range, 
lead to $r_e/r_0\approx 1.682 \ln(a/r_0)^{0.648}$ with similar accuracy. The divergence of $r_e$ as $a\rightarrow\infty$ is consistent
with the definition of $r_e$ in Ref.~\cite{verhaar1984JPA} and indicates that in 2D, although $r_{e}/a\rightarrow0$ as $a\rightarrow\infty$, 
special care might be needed while including effective range corrections in both few- and many-body approaches.

\section{Summary}

Our analysis of the long range corrections for the three-body potentials explains the origin
of the energy suppression of recombination observed in Refs.~\cite{helfrich2011PRA,ngampruetikorn2013EPL}.
The semi-analytical formulas derived here explicitly demonstrate this fact via the additional $1/|\ln(ka/\gamma)|^{2c_1}$ factor
in recombination that can be traced back to the long-range corrections of the three-body potentials.
We verify these results through numerical calculations of recombination for three identical bosons for the lowest few
values of $|M|$. We also show that recombination into deeply bound states has the same energy dependence as
recombination into weakly bound states but with a stronger suppression with increasing scattering length. This
result indicates that studies of strongly interacting ultracold 2D gases might be easier to realize than in the 3D case.
Our analysis of finite-range effects on the energies of three-boson bound states indicates that the universal regime
is approached slowly as $a$ increases. We suggest a correction term for these energies in terms 
of $r_e/a$ that gives a good description for values of $r_e/a<0.1$. 

\acknowledgments
The work was supported by AFSOR -MURI (USA).

\appendix

\section{Numerical calculations for three-body recombination} \label{Match}

Here, we give a brief description of how we calculate three-body recombination in 2D numerically,
using the hyperspherical approach. As mentioned in the main text, the first step in the calculation
is to solve the adiabatic equation, Eq.~(\ref{poteq}), to determine the three-body potentials, $U_{\nu}(R)$, 
and channel functions, $\Phi_{\nu}(R;\Omega)$. As shown in Ref.~\cite{dincao2014ARX}, Eq.~(\ref{poteq})
reduces to two coupled partial differential equations in the hyperangles $\theta$ and $\varphi$ (only one for $M=0$).
The resulting differential equations are solved by expanding $\Phi$ onto a direct product of basis splines 
in $\theta$ and $\varphi$ \cite{Splines,EsryThesis} with a proper set of boundary conditions \cite{dincao2014ARX}.
Since our goal is to calculate scattering observables at ultracold energies, we solve Eq.~(\ref{poteq}) up to
distances comparable to $R=10^5r_{\rm vdW}$ ---for scattering calculations one typically wants to solve the problem
to distances that greatly exceed the classical turning point. In order to obtain at least six digits of accuracy for
the potentials at such large distances, we used 140 basis splines for each hyperangle (for more details of our numerical 
implementation see Ref.~\cite{dincao2014ARX}). The presence of a repulsive core in the two-body potential model
used in the calculations [see Eq.~(\ref{vdWPot})] is also a factor that required us to use so many basis splines. 
In fact, to prevent the b-spline matrix elements from diverging due to this unphysical $1/r^{12}$ short-range behavior, 
we cut off the potential at very short distances.

After solving Eq.~(\ref{poteq}), we must solve the hyperradial equation in Eq.~(\ref{radeq}). 
Details of the method for solving Eq.~(\ref{radeq}) are given in Ref.~\cite{wang2011PRAb}, so we will only 
emphasize some fundamental aspects we found for three-body recombination in 2D.
The accuracy of the numerical solutions depend on various factors.
Besides the usual requirements of a dense enough hyperradial grid and enough channels included in the calculation,
the determination of scattering observables also requires Eq.~(\ref{radeq}) to be solved up to distances
where the effect of the logarithmic terms in Eq.~(\ref{URS}) are negligible compared to the first term. Only at these
distances can we math the numerical results to the asymptotic free-particle solutions
\begin{eqnarray}
f_{\nu}(R)=\left(\frac{2\mu k}{\pi}\right)^{1/2}R j_{l_\nu}(kR),\label{fsol}\\
g_{\nu}(R)=\left(\frac{2\mu k}{\pi}\right)^{1/2}R n_{l_\nu}(kR),\label{gsol}
\end{eqnarray}
where $j_l$ and $n_l$ are the regular and irregular spherical Bessel functions, respectively,
and correctly obtain the scattering observables. For recombination, $k^2=2\mu E$, 
the initial channels have $l_{\nu}=\lambda+1/2$ while the final states have $l_{\nu}=|m_{AD}|-1/2$,
as determined from the asymptotic form of the three-body potentials in Eqs.~(\ref{3BAD}) and (\ref{3BCont}), 
respectively. 

\begin{figure}[htbp]
\includegraphics[width=3.3in]{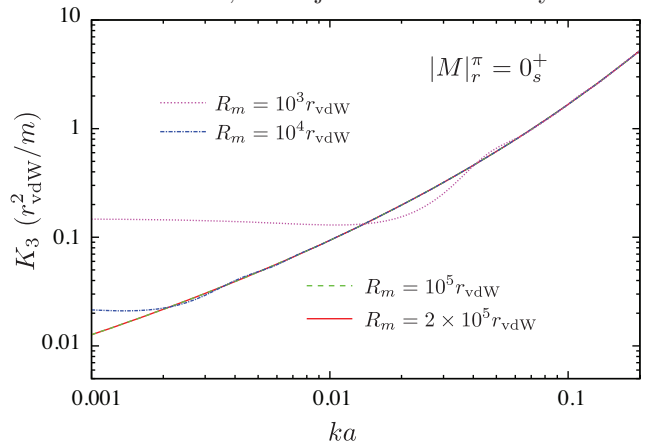}
\caption{(Color online). Low-energy behavior of three-body recombination for three identical bosons with $M^{\pi}_{r}=0^+_s$ 
with different values for $R_{m}$. Here, we adjusted the two-body interaction [Eq.~(\ref{vdWPot})] to support a single $m_{2b}=0$
bound state and produce $a=10.04r_{\rm vdW}$. For our largest value of $R_m$, the rate is converged up to three digits.}
\label{K32DRm}
\end{figure}

However, the long-range nature of the corrections to the potential we found in Eq.~(\ref{URS}) led us to pay close attention to the
convergence of $K_3$ with respect to the matching distance $R_m$. This concern arises from the fact that the spherical
Bessel functions in Eqs.~(\ref{fsol}) and (\ref{gsol}) are the solution only when all terms in Eq.~(\ref{URS}) except the first can be
neglected. Given the behavior of the corrections, it is not entirely clear that this is ever true ---but if it is, it must be at very large distances.
Rather than try to answer this formal question for these logarithmic potentials, we took the pragmatic approach of requiring convergence
with respect to $R_m$ using the asymptotic solutions we knew ---i.e., Eqs.~(\ref{fsol}) and (\ref{gsol}).

In Fig.~\ref{K32DRm}, we show the low-energy behavior for $BBB$ $M^{\pi}_{r}=0^+_s$ recombination
with different values for $R_{m}$. For this calculation, we adjusted the two-body interaction [Eq.~(\ref{vdWPot})] to support a single $m_{2b}=0$
bound state and produce $a=10.04r_{\rm vdW}$. For $R_{m}=10^3r_{\rm vdW}$, we clearly observe a change in behavior of $K_{3}$ as $ka$
approaches small values. As we increase $R_m$, the change in behavior of $K_3$ is moved towards even smaller values of $ka$. For our largest 
value of $R_m=2\times10^5r_{\rm vdW}$, no substantial difference can be noticed in $K_3$ with respect to the calculations with $R_m=10^5r_{\rm vdW}$
---the rate is converged up to three digits accuracy for the $ka$ range shown. 

\end{document}